\documentclass[aps,twocolumn, secnumarabic,nobalancelastpage,amsmath,amssymb,nofootinbib,floatfix]{revtex4-2}

\usepackage{graphicx}
\usepackage{bm}
\usepackage{bbm}
\usepackage{bbold}
\usepackage{amsmath}
\usepackage{braket}
\usepackage{color}
\usepackage[normalem]{ulem}
\usepackage{soul}

\newcommand{\cut}[1]{} 

\begin{document}

\title{Demonstration of universal control between non-interacting qubits using the Quantum Zeno effect}

\author{E. Blumenthal$^1$}
\author{C. Mor$^1$}
\author{A. A. Diringer$^1$}
\author{L. S. Martin$^2$}
\author{P. Lewalle$^{3,4}$}
\author{D. Burgarth$^5$}
\author{K. B. Whaley$^{3,4}$}
\author{S. Hacohen-Gourgy$^1$} 
\affiliation{$^1$Department of Physics, Technion - Israel Institute of Technology, Haifa 32000, Israel\\
$^2$Department of Physics, Harvard University, Cambridge, Massachusetts 02138, USA\\
$^3$Department of Chemistry, University of California, Berkeley, California 94720 USA\\
$^4$Berkeley Center for Quantum Information and Computation, Berkeley, California 94720 USA\\
$^5$Center for Engineered Quantum Systems, Dept. of Physics \& Astronomy, Macquarie University, 2109 NSW, Australia}

\begin{abstract}
The Zeno effect occurs in quantum systems when a very strong measurement is applied, which can alter the dynamics in non-trivial ways. Despite being dissipative, the dynamics stay coherent within any degenerate subspaces of the measurement. Here we show that such a measurement can turn a single-qubit operation into a two- or multi-qubit entangling gate, even in a non-interacting system. We demonstrate this gate between two effectively non-interacting transmon qubits. Our Zeno gate works by imparting a geometric phase on the system, conditioned on it lying within a particular non-local subspace. These results show how universality can be generated not only by coherent interactions as is typically employed in quantum information platforms, but also by Zeno measurements.
\end{abstract}

\maketitle

Control of quantum systems can be divided to two distinct schemes, coherent and incoherent control. Coherent control is achieved by application of control Hamiltonians to evoke deterministic time evolution. In contrast, incoherent control is based on non-deterministic measurement outcomes to prepare the system in a desired state. The two schemes may complement each other to enrich quantum control~\cite{Felix2015,martin2015deterministic,thomsen2002continuous,JacobsPurification2003,Riste2013,NKatz2008,Vool2016,SHGReview}. On the boundary between the two schemes lies the quantum Zeno effect, in which frequent measurements effectively freeze the system dynamics, holding the system at an eigenstate of the measurement observable. A more precise description shows that measurements divide the Hilbert space into subspaces with distinct eigenvalues of the measured observable, and give rise to `Zeno dynamics' within each~\cite{facchi2008QZD}. Transitions between subspaces are suppressed by measurement, but the evolution inside each subspace is completely coherent. In particular, previous work has shown that Zeno dynamics can theoretically transform a trivial (e.g., non-interacting with local control only) quantum system into one with universal control within the Zeno subspace~\cite{burgarth2014exponential} and several state entangling schemes have been proposed~\cite{wang2008quantum, shao2009onestep,zhang2011robust}. 
\begin{figure}
    \includegraphics[width=\linewidth]{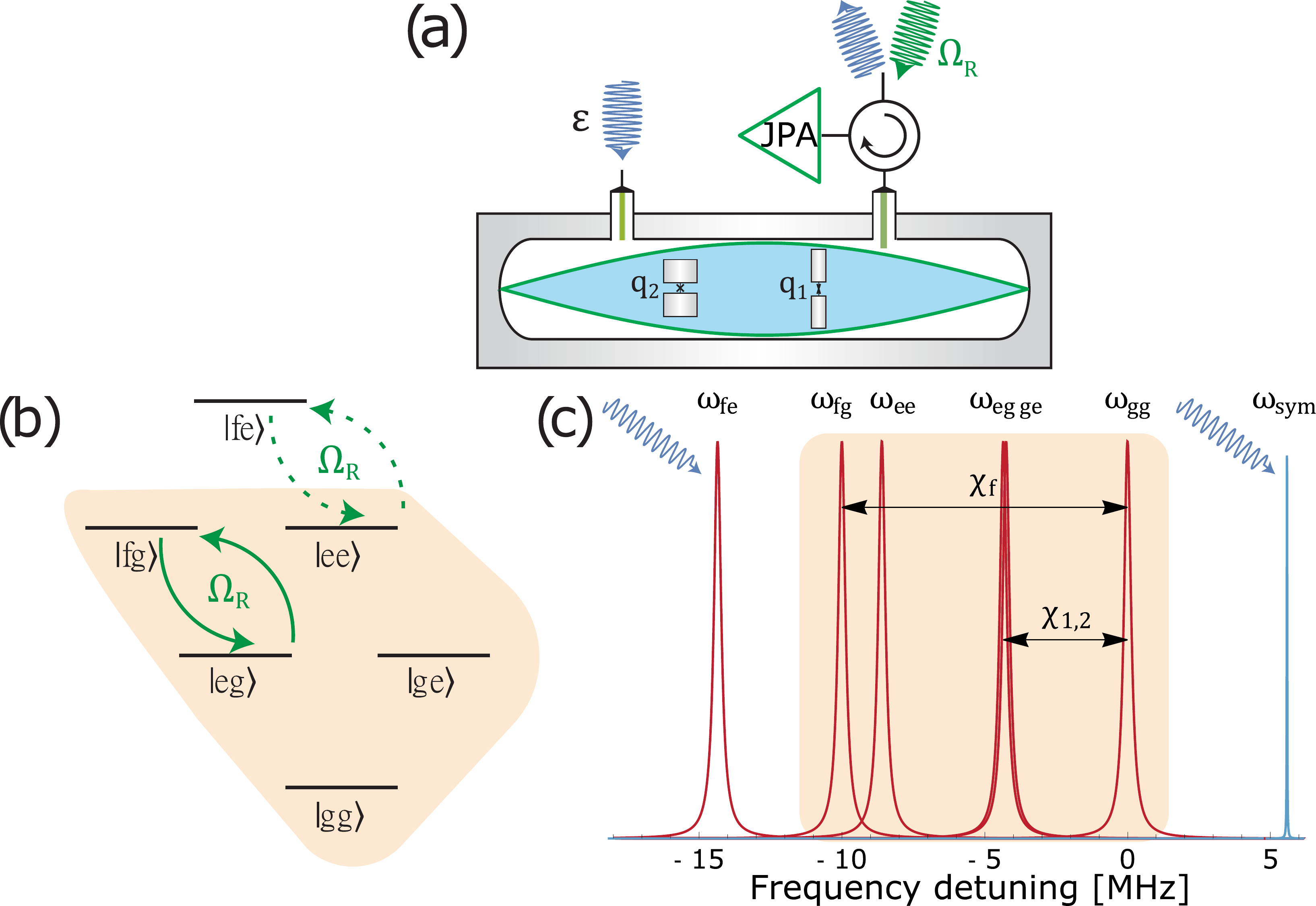}
    \caption{Experiment schematic. (a) Two transmons coupled to electromagnetic-mode of a superconducting cavity. (b) Qubit-qutrit energy level diagram, where the energy levels of each element are labeled g (ground), e (excited) and f (second excited level). The colored domain is the subspace defined by the projector $P=\mathbb{1}-\Ket{fe}\Bra{fe}$. The $\Ket{e}\leftrightarrow\Ket{f}$ transition of the qutrit q1 is Rabi driven with frequency $\Omega_R$. Dotted and solid lines are blocked and allowed transitions, respectively. (c) Cavity spectra conditioned on transmon state (red) and the applied driving tones (blue).
    }
    \label{fig:gateSetup}
\end{figure}

In this letter we show an explicit construction of such universal control, and demonstrate it in a circuit-QED system~\cite{blais2021CQED}. Our unique construction performs in a single operation, an N-Control-phase gate on N qubits, where the last qubit is required to have only one extra level, i.e., it is a qutrit. We refer to this as a Zeno gate. Specifically, we demonstrate the gate between two non-interacting transmon qubits~\cite{Koch2007}. This work is distinct from other measurement based methods that prepare entangled states~\cite{Roch2014,Riste2013,shankar2013autonomously}; the major novelty is that the dynamics here are coherent, deterministic, and allow for universality.

Technically our experimental system has a resonator induced interaction, which can yield a high fidelity gate (RIP-gate)~\cite{RIPgate,RIPgateEXP}. We actively cancel these interactions to make our system effectively non-interacting. We can then demonstrate dynamics due to the Zeno effect alone.
Our purpose is to show how universality can be switched on and off 
just by looking at a single level within a quantum system.
\begin{figure}[htp!]
    \includegraphics[width=\linewidth]{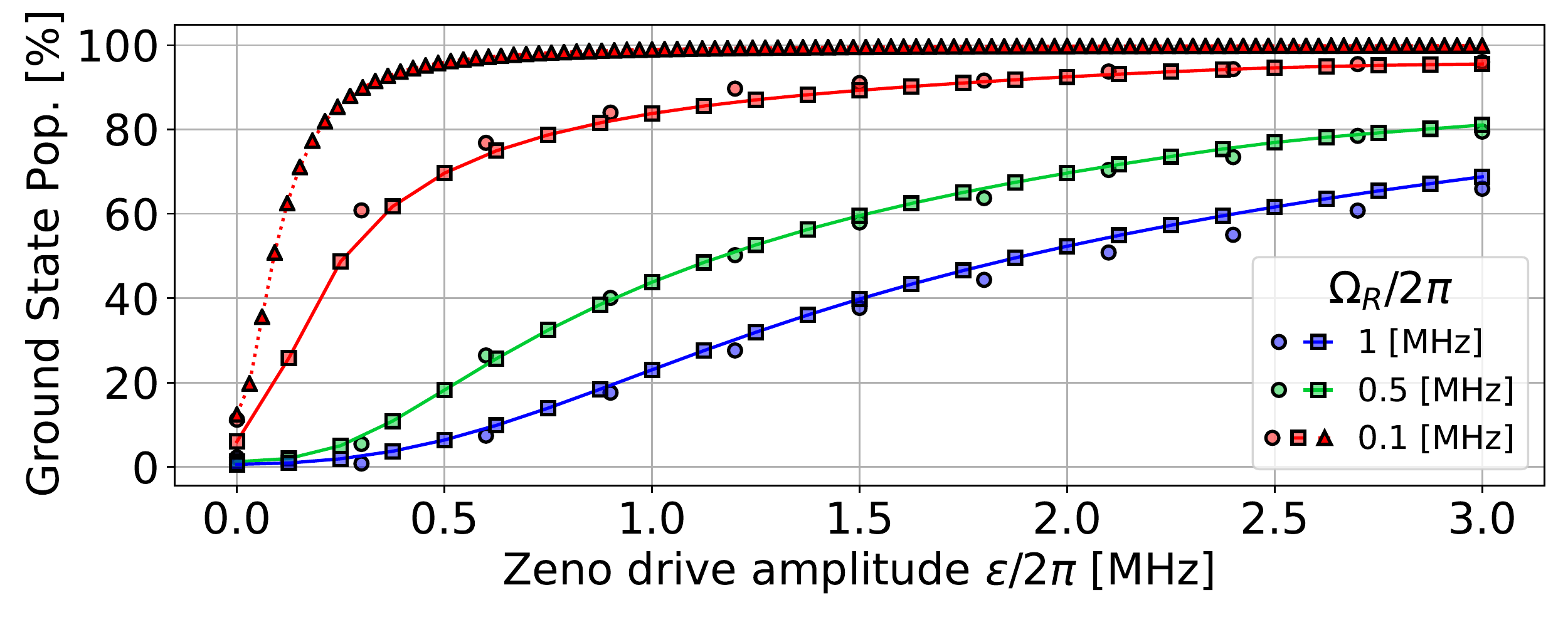}
    \caption{$\Ket{gg}$ population after starting in $\Ket{gg}$ and Rabi driving the qutrit $\Ket{g}\leftrightarrow\Ket{e}$ transition for half of an oscillation, while simultaneously Zeno driving the cavity at $\omega_{eg}$, as function of Zeno drive amplitudes $\varepsilon$. 
    Circles are experimental results, squares are numerically simulated results and triangles are an ideal simulation assuming the cavity is a Markovian bath, Eq.~\ref{eq:idealMasterEq} (the solid lines are provided to guide the eye).}
    \label{fig:zenoBlock}
\end{figure}
\begin{figure*}[htp!]
    \centering
    {\includegraphics[width=1\linewidth]{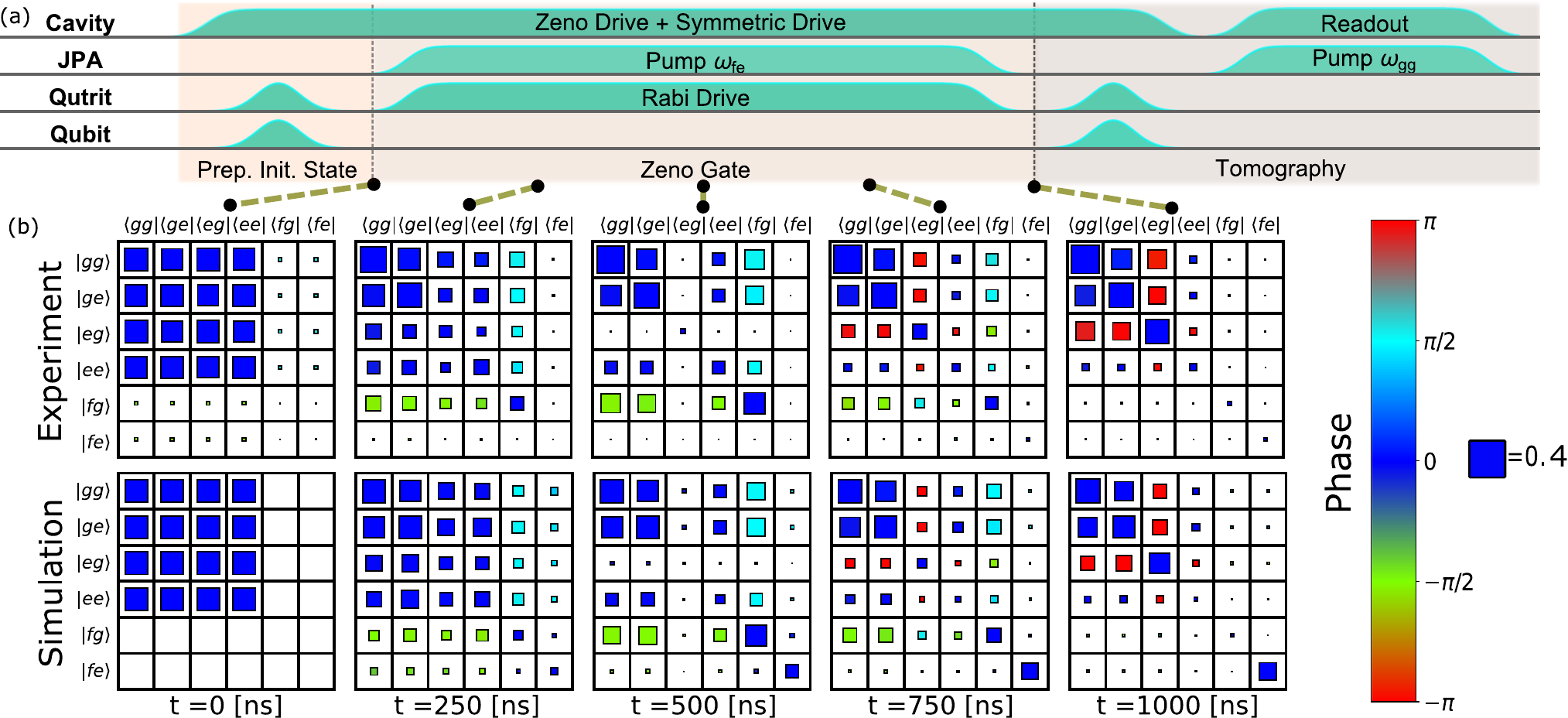}}
    \caption{(a) Pulse sequence for the Zeno dynamics. (b) The qutrit-qubit density matrix at different times with $\varepsilon / 2 \pi = 2$ MHz, starting with an initial $\Ket{++}$ state. Black squares are partially filled to represent the amplitude, where a full square stands for an amplitude of $0.4$, and the color of the filling represents the phase according to the color bar.
    Experimental results (top row), numerical simulation (bottom row).}
    \label{fig:timeEvo}
\end{figure*}
%
The Zeno dynamics we explore here rely on local operations together with non-local projections. Locally driving one transition for a full $2\pi$ rotation imparts a geometric phase of $\pi$ on the initial state. Adding rapid projections blocks transitions between the Zeno subspaces defined by the projector, and allows phase accumulation only for certain states. Choosing an appropriate non-local projector conditions the resulting phase on the state of both qubits and thereby leads to entanglement. This process is similar to entangling operations based on Rydberg blockade with neutral atoms~\cite{isenhower2010demonstration, levine2019rydberg} in the sense that a certain non-local state can not be reached by the system. The main difference is that while the Rydberg blockade is a result of strong coherent interactions~\cite{jaksch2000fast}, we use incoherent measurements to \cut{employ} perform the Zeno block.

Consider first an ideal qubit-qutrit system and infinitely rapid projections, where the qutrit $\Ket{f}$ level is an auxiliary state. We apply a Rabi drive of frequency $\Omega_R$ between $\Ket{e}$ and $\Ket{f}$, and at the same time apply rapid projective measurements of the projector $P=\mathbb{1}-\Ket{fe} \Bra{fe}$ (as depicted in Fig.\ref{fig:gateSetup}b). In the limit of infinitely rapid projections the Hamiltonian reads \cite{facchi2008QZD}
\begin{equation}
    \begin{split}
        &H_{\textrm{Zeno}} = PHP \\ &=i\hbar\dfrac{\Omega_R}{2}P\left(\Ket{e} \Bra{f} - \Ket{f} \Bra{e}\right)\otimes \left(\Ket{g} \Bra{g} + \Ket{e} \Bra{e}\right) P \\
        &= i\hbar\dfrac{\Omega_R}{2}(\Ket{eg} \Bra{fg} - \Ket{fg} \Bra{eg})
    \end{split}
    \label{idealHamilt}
\end{equation}
where $H = \tfrac{1}{2}i\hbar\,\Omega_R\left(\Ket{e} \Bra{f} - \Ket{f} \Bra{e}\right)\otimes \mathbb{1} $ is the Rabi oscillation Hamiltonian without projections.
The $\Ket{eg}\leftrightarrow\Ket{fg}$ transition is allowed and the $\Ket{ee}\leftrightarrow\Ket{fe}$ transition is blocked and does not appear in Eq.~\ref{eq:idealMasterEq}, as shown in Fig.~\ref{fig:gateSetup}b with solid and dotted arrows, respectively.
Assuming the system started in the subspace defined by $P$, it will remain there and undergo coherent evolution governed by $U_{\mathrm{Zeno}}=\exp(-iH_{\mathrm{Zeno}}t/\hbar)$. Applying the operation for a time $t=2\pi /\Omega_R$, one full oscillation, the $\Ket{eg}$ state acquires a $\pi$ phase. Thus, our operation is equivalent to a Control-phase gate up to local operations. 
This scheme can be expanded to entangle multiple qubits and one qutrit by measuring the projector $P=\mathbb{1}-\Ket{fee...e}\Bra{fee...e}$. A $\pi$ phase will be acquired by states $\Ket{exx..x}$, except for $\Ket{ee..e}$, where $x\in [e,g]$. This operation is equivalent to a N-Control-phase gate.

The key experimental requirement is the ability to apply the projector $P$. In a realistic setup, the projection application rate is not infinite, and the system may be described either by a sequence of projections with a finite time interval between them, or by a continuous measurement~\cite{facchi2008QZD}. We focus on the latter case as it fits our experimental circuit-QED scheme. Continuous measurements of the projector $P$ at a rate of $\Gamma$ can be modeled by the master equation
\begin{align}
    \frac{d\rho}{dt}=-i[H,\rho]+\Gamma\mathcal{D}[P]\rho
    \label{eq:idealMasterEq}
\end{align}
where $\mathcal{D}[\cdot]$ is the standard Lindblad dissipator that models coupling to a Markovian bath. The finite measurement rate introduces a chance for the system to escape the Zeno subspace. The corresponding gate error in diamond norm \cite{watrous} can be bounded as $\mathcal{E}_\diamond < 38\;\Omega_R/\Gamma$~\cite{Suppmat}.

Eq. \ref{eq:idealMasterEq} describes the system in the Markovian regime where the bath ``loses its memory'' faster than the system evolution rate. This timescale puts an upper bound on the Rabi frequency $\Omega_R$. Beyond this frequency, in the non-Markovian regime, the system cannot be described by the simple form of Eq. \ref{eq:idealMasterEq}. In our system this time scale is given by the cavity linewidth $\kappa$. However, to maximize our gate fidelity, we perform the gate at a rate faster than the system decoherence, and show that Zeno dynamics are qualitatively the same, differing only in showing a limited blocking ability. This is in line with a recently predicted unification of Zeno physics arising through a wide range of mechanisms \cite{unity1,unity2}.

We implement the Zeno gate on a circuit-QED system composed of two transmons \cite{Koch2007} dispersively coupled to a superconducting 3D cavity, Fig.~\ref{fig:gateSetup}a. The system was designed to optimize implementation of the non-local measurement $P$, while minimizing qubit-qutrit interactions. The transmons were fabricated with far detuned transition frequencies of $\omega_{q1}/2\pi=3.28$ GHz, $\omega_{q2}/2\pi=6.24$ GHz and anharmonicities of $\alpha_1/2\pi=-175$ MHz and $\alpha_2/2\pi=-225$ MHz respectively. We use $q1$ as the qutrit. The cavity mode frequency was $\omega_c/2\pi=7.32$ GHz. The linewidth $\kappa/2\pi = 0.15$ MHz, was predominantly set by the strongly coupled port. The transmons-cavity dispersive couplings were $\chi_1/2\pi=-4.25$ MHz, $\chi_2/2\pi=-4.35$ MHz and the $\Ket{f}$ state was $\chi_f/2\pi=-10$ MHz.
The system--cavity interaction is well described by the dispersive Hamiltonian in the interaction picture \cite{blais2021CQED}
\begin{equation}
    \begin{split}
        H_{\text{disp}}/\hbar&=\left(\chi_1\Ket{e_1}\Bra{e_1} + \chi_2\Ket{e_2}\Bra{e_2} +\chi_f\Ket{f}\Bra{f}\right)a^\dagger a\\
        &+\alpha_1\Ket{f}\Bra{f},
    \end{split}
    \label{eq:dispHamilt}
\end{equation}
where $a^\dagger$ and $a$ are the creation and annihilation operators of photons in the cavity, and subscripts in the kets label the qubits. We omitted the residual direct qubit-qutrit interaction, which was measured using Ramsey interferometry, between the $\Ket{ge}$ and $\Ket{ee}$ states. We measured $~30$ KHz, negligible for the timescales of our experiment.

Eq. \ref{eq:dispHamilt} shows that the cavity acquires a frequency shift that depends on the qubit-qutrit state. In the $\chi\gg\kappa$ regime, the cavity resonance frequencies for each state of the qubits are well separated, Fig.~\ref{fig:gateSetup}c. Probing the cavity resonance frequency allows us to deduce the qubit-qutrit state. We do this by driving the cavity through the weakly coupled port, and monitoring the output through the strongly coupled port. We continuously measure the projector $P$ by driving the cavity at a frequency of $\omega_{fe} = \omega_{gg}+\chi_f+\chi_2$, which is the resonance frequency when the system is in $\Ket{fe}$. We refer to such a measurement as a ``Zeno drive". \cut{In the $\chi\gg\kappa$ regime, where the cavity resonance frequencies for each state of the qubits are well separated (see Fig.~\ref{fig:gateSetup}c), this choice makes the transition from $\Ket{ee}$ to $\Ket{fe}$ much more strongly measureable than any other transition, allowing approximation of the desired Zeno projector.} The output signal is amplified using a flux-pumped Josephson Parametric Amplifier (JPA), with design as in~\cite{hacohen2016dynamics}. Changing the pumping frequency, we sequentially amplify signals of different frequencies. We amplify the Zeno drive signal at $\Omega_{fe}$ first, followed by the readout signal at $\Omega_{gg}$. The former enables us to detect whether the system escaped the Zeno subspace during the gate operation, the latter is used for tomography. We note that for the Zeno block to occur, the measurement may be performed by the ``environment". High quantum efficiency is not required to implement the gate and is not even necessary to observe the measurement outcome. However, this is important for high fidelity post-selection.

Before proceeding to the entangling dynamics, we first characterize the Zeno block probability as function of the drive amplitude $\varepsilon$. We demonstrate this here on the two lowest states of the qutrit $q1$. We apply a Zeno drive at $\omega_{eg}$ and Rabi drive the transition for $t=\pi/\Omega_R$. We measure the probability to stay in $|gg\rangle$, as function of the Zeno drive amplitude for three different Rabi frequencies, see Fig.~\ref{fig:zenoBlock}. This procedure resembles that in \cite{ZenoSlichter}, with slight differences because that experiment was conducted using a quantum trajectory approach in the steady state. Furthermore, Ref.~\cite{ZenoSlichter} operated in the $\Omega_R < \kappa$ regime, meaning the cavity could be modeled as a Markovian bath and the textbook jump rate value of $P_{\text{jump}}=\Omega_R^2/2 \Gamma$~\cite{Misra1977,facchi2008QZD} was observed. Here we show that even beyond this regime, the Zeno effect still blocks, albeit with a reduced effectiveness. 

Fig.~\ref{fig:zenoBlock} shows the expected qualitative behaviour where the blocking probability increases with the drive amplitude, and decreases with increasing Rabi frequency. Quantitatively the data agree with the numerical simulation of the master equation of the full qubit-qutrit-cavity system~\cite{Suppmat}. However, our system can be simplified to Eq. \ref{eq:idealMasterEq} only in the limit $\Omega_R \ll \kappa$. In that limit, $\Gamma = 4\varepsilon^2 / \kappa$ \cite{Gambetta2008}. Even at $\Omega_R/2 \pi=0.1~\mathrm{MHz} =  2\kappa/3$ we can still see a deviation from Eq.\ref{eq:idealMasterEq} (red symbols), with a reduced blocking probability relative to that expected for this value of $\Gamma$. Recent experiments have probed pertinent regimes in greater detail \cite{Szombati2020, koolstra2021monitoring}, and begun to illustrate how lag in the cavity state ``following'' a qubit on a timescale $\kappa^{-1}$ impacts subsequent measurement mediated by the cavity.
While a slower Rabi frequency is better in terms of realizing the Zeno effect, our gate time needs to be significantly shorter than the system coherence times ($T_1^{e\rightarrow g}=52$ $\mu$s, $T_1^{f\rightarrow e}=12.9$ $\mu$s, $T_2^{* e\leftrightarrow g}=22.2$ $\mu$s, $T_2^{* f\leftrightarrow e}=5.8$ $\mu$s for the qutrit, and $T_1=18.9$ $\mu$s, $T_2^*=15.7$ $\mu$s for the qubit). We set $\Omega_R/2 \pi=1$ MHz (blue in Fig.~\ref{fig:zenoBlock}).

Until now we have discussed only the effect of the Zeno drive on the transition that we wish to block.
However, residual effects on the rest of the states also emerge. Due to the non-zero cavity linewidth, driving at $\omega_{fe}$ will create a small coherent displacement even if the system is not in $\Ket{fe}$. In the frame rotating with the drive frequency, at a \textit{steady state} this coherent state is $\alpha_{ij} = \dfrac{\varepsilon}{i\Delta_{ij}+\kappa/2}$, where $\varepsilon$ is the drive amplitude, and $\Delta_{ij}$ is the detuning between the cavity resonance frequency $\omega_{ij}$ and the drive frequency when the qubits are in $\Ket{ij}$. In our $\chi\gg\kappa$ regime, we can write $\rho_{ij,k\ell}(t)=e^{i\mu_{ij,k\ell}t}\rho_{ij,k\ell}(0)$, where $\mu_{ij,k\ell}=(\omega_{ij}-\omega_{k\ell})\alpha_{k\ell}^*\alpha_{ij}$ such that a phase will be acquired between each pair of states at a rate of $\mathrm{Re}[{\mu}_{ij,k\ell}]=\dfrac{(\omega_{ij}-\omega_{k\ell})|\varepsilon|^2}{\Delta_{ij}\Delta_{k\ell}}$ and coherence will be lost due to measurement-induced dephasing at a rate of $\mathrm{Im}[{\mu}_{ij,k\ell}]=\dfrac{(\omega_{ij}-\omega_{k\ell})^2|\varepsilon|^2\kappa}{2\Delta_{ij}^2\Delta_{k\ell}^2}$. This is the RIP-gate, where conditional phase accumulation leads to entanglement of the qubits \cite{RIPgate, RIPgateEXP}. To demonstrate the entanglement caused \textit{only} by Zeno dynamics, we negate this effect by applying an additional drive. It is applied to the cavity, at a frequency that is symmetric to the Zeno measurement drive frequency with respect to $\omega_{eg}$ and $\omega_{ge}$, so that $\omega_{sym}=\omega_c+(\chi_f+\chi_2)-(\chi_1+\chi_2)=\omega_c+\chi_f-\chi_1$, as depicted in Fig.~\ref{fig:gateSetup}. This symmetric drive balances the phase accumulation, such that this no longer generates entanglement. We note that while the phase accumulation is given above for the steady state, a cancellation of the phase by the symmetric drive should also occur in the transient regime. We confirmed this by numerical simulation as well as by Ramsey interferometry between $\Ket{gg}\leftrightarrow \Ket{eg}$ and $\Ket{ge}\leftrightarrow \Ket{ee}$ while applying both the Zeno and the symmetric drives to the cavity~\cite{Suppmat}.
\begin{figure}[htp!]
    \centering
    \includegraphics[width=1\linewidth]{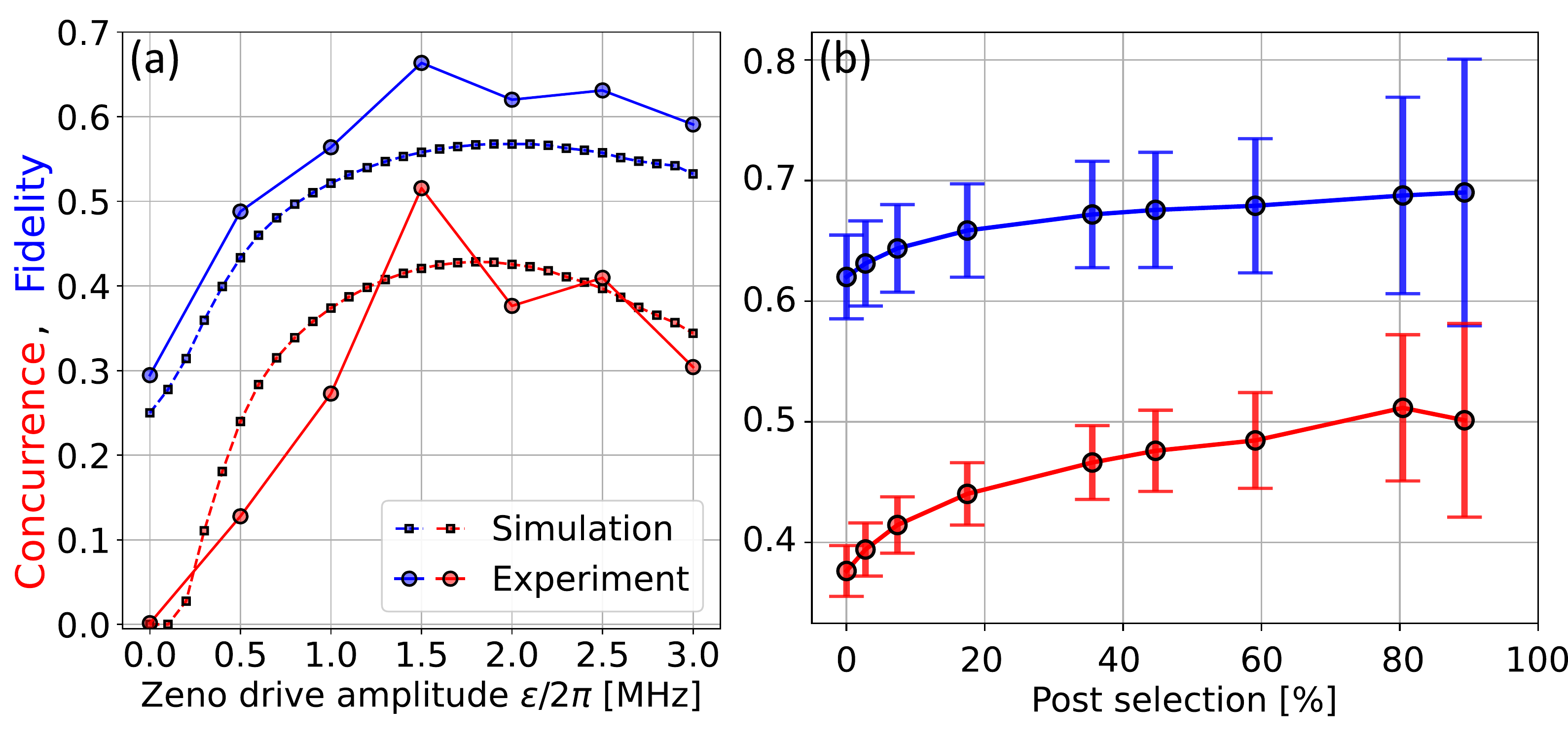}
    \caption{Gate fidelity (blue) and concurrence (red) versus the amplitude of the Zeno drive (a), and as function of the post-selection percentage for a Zeno drive amplitude of $2$ MHz (b). The fidelity is calculated with respect to an ideal state, obtained by applying $\mathbb{1}-2\Ket{eg}\Bra{eg}$ to our initial state $\Ket{++}$. Circles are experimental results and squares are numerical results (lines are guide to the eye).}
    \label{fig:fdltyVSEpsilon}
\end{figure}
The driven system Hamiltonian with both the Zeno drive and the symmetric drive, in the frame rotating at $\omega_{gg}$ reads
\begin{equation}
    \begin{split}
        H_{\text{driven}}/\hbar = H_{\text{disp}}/\hbar &+ i\varepsilon\left(ae^{-i(\chi_f+\chi_2)t}-a^\dagger e^{i(\chi_f+\chi_2)t}\right) \\
        &+ i\varepsilon\left(ae^{-i(\chi_2-\chi_f)t}-a^\dagger e^{i(\chi_2-\chi_f)t}\right).
        \label{eq:fullHamilt}
\end{split}
\end{equation}
To perform the Zeno gate we turn on the above drives and then initialize the system in the $(\Ket{g_1}+\Ket{e_1})(\Ket{g_2}+\Ket{e_2})$ state, see Fig.~\ref{fig:timeEvo}a (for other initial states see~\cite{Suppmat}). We then apply the Rabi drive $\Ket{e_1}\leftrightarrow\Ket{f}$ at the Stark shifted frequency for a time of $2\pi/\Omega_R$. Finally, we apply a set of tomography pulses~\cite{Suppmat}, and apply a readout pulse. 

We sample the time evolution of the system, as shown in Fig.~\ref{fig:timeEvo}b. We see that the final state, after $1$ $\mu$s, is entangled since $\Ket{eg}$ has acquired a phase of $\pi$. The main discrepancy between the experiment and simulation is the population of $\Ket{fe}$, which is much smaller in the experiment than in the simulation. This is most likely due to the Zeno drive populating the cavity with a large coherent state once an escape occurs, thus shifting the qutrit resonance frequency and preventing the tomography pulse from correctly mapping $\Ket{fe}$~\cite{Suppmat}. The lost $\Ket{fe}$ population is then translated to a completely mixed state, therefore increasing the computational subspace population, which can cause a calculated fidelity increase, as discussed below.
In addition, the relaxation rate may be increased during the gate due to the large Zeno drive amplitude~\cite{DDephasingSlichter, sank2016dressed, hanai2021intrinsic, Lescanne2019escape}. 

We performed this procedure with varying Zeno drive amplitudes and calculated the fidelity and concurrence of the final state, as shown in Fig.~\ref{fig:fdltyVSEpsilon}a. Since we start with the state $(\Ket{g_1}+\Ket{e_1})(\Ket{g_2}+\Ket{e_2})$ it is reasonable to use the fidelity of the final state as a proxy for the gate fidelity. As a check, we applied the gate to other initial states, obtaining similar quality results~\cite{Suppmat}. Concurrence is a measure of entanglement between qubits that is non-zero only for entangled states~\cite{Wooters1998Entanglement}. We calculate this on the states in the computational subspace. Increasing the Zeno drive amplitude increases the measurement rate, leading to a higher blocking probability and therefore higher fidelity and concurrence; on the other hand, this also leads to an increased dephasing rate $\mathrm{Im}[\mu_{ij,k\ell}]$, due to the finite $\kappa$. This causes the reduced fidelity and concurrence observed at higher drive amplitudes $\varepsilon$. Furthermore, we can see that the experimental results consistently achieve higher fidelity than the simulated results, while the experimental concurrence does not. This small discrepancy in gate fidelity between the experimental results and the numerical simulation is caused primarily by the incorrect mapping of $\Ket{fe}$ in the tomography process, as explained above. 

The main source of infidelity for the gate is escapes from the Zeno subspace, which can be detected using the JPA. This capability allows us to perform the gate probabilistically but with a higher chance of success by post-selecting on the JPA signal. To demonstrate this, we post-selected our tomography results based on the amplitude of the transmitted signal, for the case of $\varepsilon/2 \pi=2$ MHz. Fig.~\ref{fig:fdltyVSEpsilon}b shows an increase in both gate fidelity and concurrence with the post-selection percentage. The increase is limited by the fidelity of our error-detection, which was $\sim75\%$ although our single-shot readout fidelity was $\sim93\%$, due to the measurement time being limited by the gate time and by the increased relaxation rate from the state $\Ket{fe}$.


We have presented a system where universal control was turned on by a Zeno measurement alone. Although the measurement acts trivially in the computational subspace, it nevertheless has a non-trivial effect on the dynamics within that subspace. To demonstrate universality we performed an explicit gate on 2 qubits. The concept can be extended and works simultaneously on multiple-qubits. 

To create an effectively non-interacting system and observe dynamics due to Zeno alone, we actively cancelled the RIP-gate mechanism. In our system the RIP-gate alone would yield better performance for computational purposes. However if we consider a hypothetical system with no interactions between the qubits (possibly different type of qubits) and where the measurement drive performs only the measurement with no additional entangling effect, then the Zeno will truly be the only coherent control mechanism. 

Overall this experiment 
emphasizes the ability of the Zeno effect to turn the trivial dynamics of an apparently non-interacting system into universal control, providing proof-of-concept for an entirely novel control strategy. 

\begin{acknowledgments}
    This Research was supported by ISF grant No. 1113/19, US-Israel BSF grant No. 2020166, and Technion's Helen Diller Quantum Center. P.L. and K.B.W. were partially supported by the U.S. Department of Energy, Office of Science, National Quantum Information Science Research Centers, Quantum Systems Accelerator. D.B. acknowledges funding by the ARC (project numbers FT190100106, DP210101367, CE170100009).
\end{acknowledgments}

\section*{Contributions}
D.B., L.S.M and S.H.-G. conceived the study.
The device was fabricated by C.M. and A.A.D.
E.B. and A.A.D. constructed the experimental setup. 
The experiment and data analysis was done by E.B., assisted by A.A.D.
Theoretical modelling was done by L.S.M., S.H.-G., D.B., E.B., and P.L.
E.B. and S.H.-G wrote the manuscript.
All authors contributed to discussions and preparation of the manuscript.
All work was carried out under the supervision of S.H.-G and K.B.W.

\bibliographystyle{apsrev4-2}
\bibliography{bib}

\pagebreak
\clearpage
\widetext
\begin{center}
\textbf{\large Demonstration of an entangling gate between non-interacting qubits using the Quantum Zeno effect - Supplementary Information}
\end{center}
\setcounter{equation}{0}
\setcounter{figure}{0}
\setcounter{table}{0}
\setcounter{page}{1}
\makeatletter
\renewcommand{\theequation}{S\arabic{equation}}
\renewcommand{\thefigure}{S\arabic{figure}}
\renewcommand{\bibnumfmt}[1]{[S#1]}
\renewcommand{\citenumfont}[1]{S#1}

\onecolumngrid
\subsection*{Device parameters}
The superconducting 3D cavity was made of tin plated copper, and sealed with indium. The cavity supported a $\text{TEM}_{101}$ mode of $\omega_c /  2\pi  = $7.32GHz. The transition frequencies of the transmons were far detuned from each other and from the cavity mode in order to achieve dispersive coupling and suppress $2^{nd}$-order interactions (through the cavity mode) between the transmons. In order for the dispersive coupling constant $\chi$ to be roughly equal for both transmons, $q_1$ was fabricated with longer pads compared with $q_2$. Thus, setting the dipole coupling $g_1\approx430$ MHz for $q_1$ and $g_2\approx110$ MHz for $q_2$. The transmons and JPAs were fabricated by Aluminum deposition on resist patterns formed by electron beam lithography with a layer of ZEON ZEP 520A resist on a layer of MicroChem 8.5 MMA EL11 resist on top of a silicon substrate. Development of the resist was done in room temperature for MMA and at 0C for ZEP. The Al/AlOx/Al Josephson junctions were fabricated using a suspended bridge fabrication process \cite{BridgeJJs} for the JPAs and a bridge-free process \cite{PlusMethod} for the transmons. 

We used a JPA in phase-sensitive mode to amplify the Zeno drive signal of frequency $\omega_{fe}$ by $~12$ dB and our readout signal of frequency $\omega_{gg}$ by $~15$ dB. The frequencies are separated by $~14$ MHz, and amplifying was enabled by changing the flux-pump frequency. The pumps were applied to the system sequentially, with a $256$ ns delay between them, due to the JPA finite bandwidth. The JPA had a $~3.6$ MHz bandwidth, corresponding to a single photon decay rate of $1/\kappa_{JPA}\approx50$ ns.

\begin{figure*}[htp!]
    \centering
    \includegraphics[width = 0.5\linewidth]{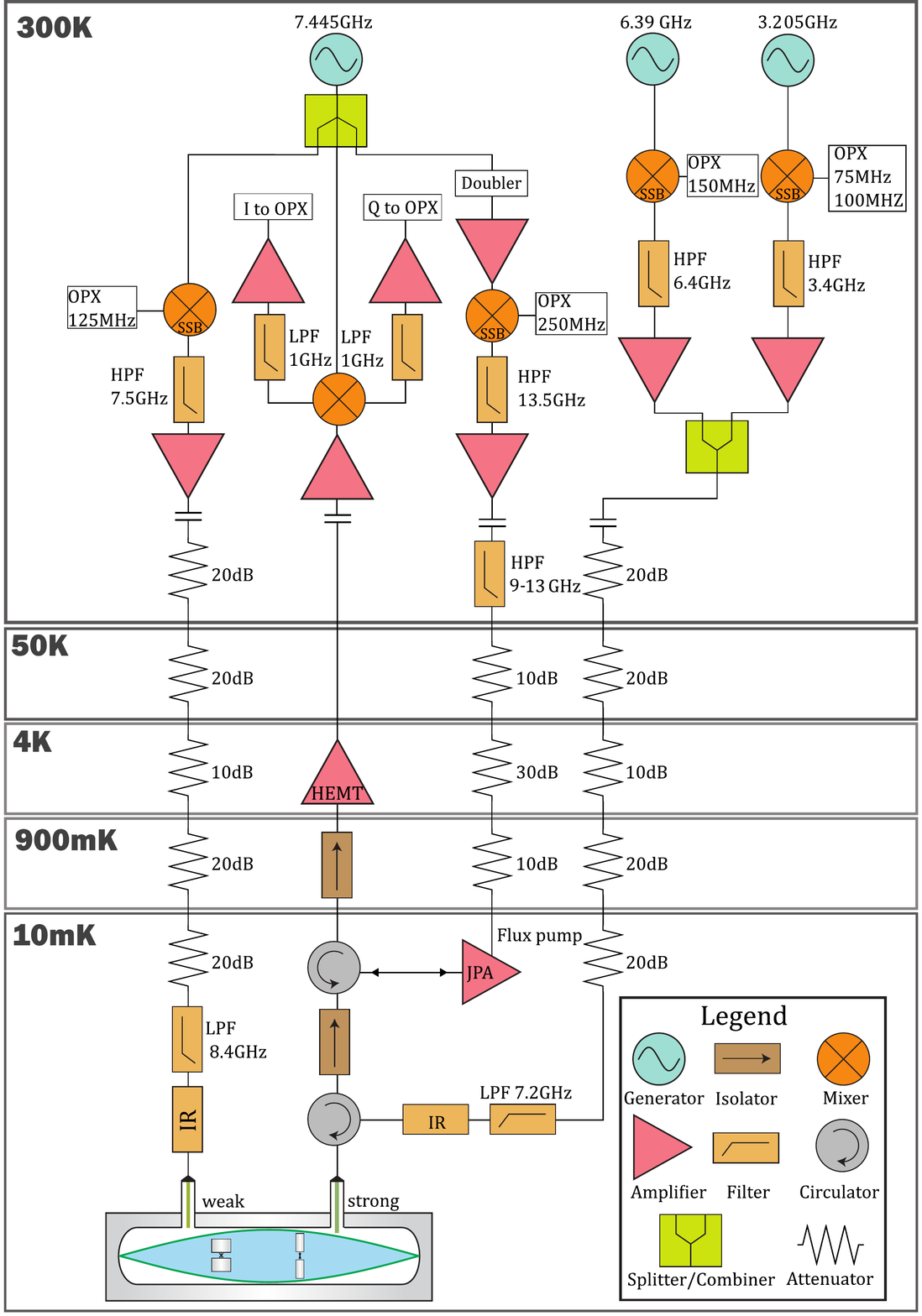}
    \caption{The schematics of the full experimental system. We apply all the drives to our system using modulated signals, using single sideband modulators, which are controlled with Operator-X (OPX) by Quantum Machines. We use one signal generator to drive both qutrit transitions. We set it at a frequency of $100$ MHz above the $\Ket{g}\leftrightarrow\Ket{e}$ transition frequency and $75$ MHz below the $\Ket{e}\leftrightarrow\Ket{f}$ transition frequency, so we can generate these intermediate frequencies with the OPX. We use another signal generator to drive the qubit. The signal generator is set to $150$ MHz above the qubit's transition frequency. We use a third signal generator set to $125$ MHz above the cavity frequency, to drive the cavity, pump the JPA and down-convert the readout signal after it passes through the cavity. The pumping signal is frequency multiplied to drive the JPA at the doubled frequency of the cavity while maintaining the phase between the cavity drive and the pump.
    }
    \label{fig:fullSystemSchematics}
\end{figure*}

The full system schematics are describe in Fig.~\ref{fig:fullSystemSchematics}.

\subsection*{Escape chance due to finite measurement rate}
Let us consider a GKLS equation \begin{equation}
		\frac{d\rho }{dt} =-i\mathcal{H}\rho+\Gamma \mathcal{D}[P]\rho
	\end{equation}
	where $i\mathcal{H}\rho\equiv -i[H,\rho]$ where for the time being $H=H^\dagger$ and $P^\dagger P =P$ arbitrary. The dissipator is explicitly given by
	\begin{equation}
		\mathcal{D}[P]\rho\equiv P\rho P^\dagger - (P^\dagger P \rho+\rho P^\dagger P )/2=P\rho P-(P\rho +\rho P)/2
	\end{equation}
	Using the notation of \cite{unity1}, $\mathcal{D}$ is already in spectral representation as $\mathcal{D}=d_0 \mathcal{P}_0+d_1 \mathcal{P}_1$ with eigenvalues $d_0=0,d_1=-\frac{1}{2}$ and spectral projectors $\mathcal{P}_1=-2\mathcal{D}$ and  $\mathcal{P}_0=\textrm{id}-\mathcal{P}_1$. $\mathcal{P}_0$ is the only peripheral part (e.g., corresponding eigenvalue on the imaginary axis), so $\mathcal{P}_\varphi =\mathcal{P}_0.$ The corresponding reduced resolvent is $S_0=-2\mathcal{P}_1$ and the Zeno generator $\mathcal{H}_Z=\mathcal{P}_0\mathcal{H}\mathcal{P}_0$.
We are interested in the distance between the full evolution and the Zeno evolution. Let us first look at this on the peripheral (non-decaying) subspace, using Eq. (B.13) of \cite{unity1},
\begin{equation}
\label{bound1}
	e^{t(\Gamma \mathcal{D}-i\mathcal{H})}\mathcal{P}_0-e^{t(\Gamma \mathcal{D}-i\mathcal{H}_Z)}\mathcal{P}_0=
	\frac{1}{\Gamma} \left( iS_0\mathcal{H}\mathcal{P}_0e^{-it \mathcal{H}_z}-ie^{t(\Gamma \mathcal{D}-i\mathcal{H})}S_0\mathcal{H}\mathcal{P}_0+ 
	 \int_0^t ds e^{(t-s)(\Gamma \mathcal{D}-i\mathcal{H})}[\mathcal{H},S_0\mathcal{H}\mathcal{P}_0]\mathcal{P}_0e^{-is \mathcal{H}_z} \right)
\end{equation}
It is both convenient and operationally meaningful to bound this expression in the diamond norm $\|\cdot\|_\diamond$  in which CPTP maps are contractions \cite{watrous}. $e^{t(\Gamma \mathcal{D}-i\mathcal{H})}$, $\mathcal{P}_0 e^{-it \mathcal{H}_z}$ and $\mathcal{P}_\varphi=\mathcal{P}_0$ are automatically CPTP. Note that $\mathcal{P}_1$ is not necessarily CPTP, but we have $\|\mathcal{P}_1\|_\diamond=\|\textrm{id}-\mathcal{P}_0\|_\diamond\le 2.$ We can therefore bound the right hand side of Eq. (\ref{bound1}) by
\begin{equation}
	\left\|\left(e^{t(\Gamma \mathcal{D}-i\mathcal{H})}-e^{t(\Gamma \mathcal{D}-i\mathcal{H}_Z)}\right)\mathcal{P}_0\right\|_\diamond \le\frac{8}{\Gamma}(\|\mathcal{H}\|_\diamond+t \|\mathcal{H}\|_\diamond^2)
\end{equation}
where we used the triangle inequality and sub-multiplicity of the diamond norm. Let us now look at the decaying part $\mathcal{P}_1$. We can compute it using an integral representation as
\begin{equation}\label{bound2}
	\left( e^{t(\Gamma \mathcal{D}-i\mathcal{H})}-e^{t(\Gamma \mathcal{D}-i\mathcal{H}_Z)}\right)\mathcal{P}_1=
	-i\int _0^t
ds e^{(t-s)(\Gamma \mathcal{D}-i\mathcal{H})}\mathcal{H}e^{-s\Gamma /2}\mathcal{P}_1
\end{equation}
This is bounded as
\begin{equation}
\left\|	\left( e^{t(\Gamma \mathcal{D}-i\mathcal{H})}-e^{t(\Gamma \mathcal{D}-i\mathcal{H}_Z)}\right)\mathcal{P}_1\right\|_\diamond \le 4\|\mathcal{H}\|_\diamond\frac{1-e^{-t\Gamma/2}}{\Gamma}
\end{equation}
We can furthermore bound $\|\mathcal{H}\|_\diamond\le 2\|H\|_\infty$. In total, using $\mathcal{P}_0+\mathcal{P}_1=\textrm{id}$ we obtain
\begin{equation}\label{bound22}
    \|e^{t(\Gamma \textbf{}-i\mathcal{H})}-e^{t(\Gamma \mathcal{D}-i\mathcal{H}_z)}\|_\diamond\le \frac{16\|\mathcal{H}\|_\infty}{\Gamma}\left(1+t\|\mathcal{H}\|_\infty +\frac{1-e^{-t\Gamma/2}}{2}\right)
\end{equation}
Since $\|e^{t(\gamma \mathcal{D}-i\mathcal{H}))}\mathcal{P}_1\|_\diamond \le e^{-t\Gamma/2}$, we can finally write 
\begin{equation}\label{bound33}
    \mathcal{E}_\diamond \equiv\|e^{t(\Gamma \mathcal{D}-i\mathcal{H})}-e^{-it\mathcal{H}_z}\mathcal{P}_0\|_\diamond\le\frac{16\|\mathcal{H}\|_\infty}{\Gamma}\left(1+t\|\mathcal{H}\|_\infty +\frac{1-e^{-t\Gamma/2}}{2}\right)+e^{-t\Gamma/2}
\end{equation}
This bound is completely general and improves the bounds given in \cite{unity1} through a better choice of norm. In our experiment, $\|H\|_\infty =\frac{\Omega_R}{2}$ and gate time $t=\frac{2\pi}{\Omega_R}$, and we obtain 

\begin{equation}
	\mathcal{E}_\diamond\le \frac{8 \Omega_R}{\Gamma}(1+\pi +\frac{1-e^{-\pi\Gamma/\Omega_R }}{2})+e^{-\pi\Gamma/\Omega_R }
\end{equation}
This bound becomes non-trivial (that is, smaller $2$) only when $\frac{\Omega_R}{\Gamma }<0.06,$ and can be linearly loosened to obtain $\mathcal{E}_\diamond \le 38\frac{\Omega_R}{\Gamma}$ as mentioned in the main text.
\subsection*{Simulation details}

We simulated the full experimental system, including the transmons, the cavity and their interactions under the dispersive approximation, using the master equation
\begin{equation}
    \dot{\rho}=-i[H,\rho] +\kappa\mathcal{D}[a]\rho +\sum_{i} \Gamma_i\mathcal{D}[\sigma_-^{(i)}]\rho +\sum_{i} \dfrac{\gamma_i}{2}\mathcal{D}[\sigma_z^{(i)}]\rho
\end{equation}
where $\Gamma_i$, $\gamma_i$, $\sigma_-^{(i)}$ and $\sigma_z^{(i)}$ are the relaxation rate, dephasing rate, lowering operator and Pauli-z operator of the $i^{\text{th}}$ transition, respectively. The transitions are $i\in[f\leftrightarrow e_1, e_1\leftrightarrow g_1$, $e_2\leftrightarrow g_2$]. The master equation takes into account the natural dephasing and relaxation rate of the qutrit, the qubit and the cavity. The full Hamiltonian we used written in the interaction picture is
\begin{equation}
    \begin{split}
        H/\hbar&=\left(\chi_1\Ket{e_1}\Bra{e_1} + \chi_2\Ket{e_2}\Bra{e_2} +\chi_f\Ket{f}\Bra{f}\right)a^\dagger a\\
        &+\dfrac{\alpha_c}{2}a^\dagger aa^\dagger a +\alpha_1\Ket{f}\Bra{f}\\
        &+i\dfrac{\Omega_R}{2}\left(e^{-i(\alpha_1+\delta_{ef}(\varepsilon))t}\Ket{e_1}\Bra{f}-e^{i(\alpha_1+\delta_{ef}(\varepsilon))t}\Ket{f}\Bra{e_1}\right) \\
        & + \frac{\delta _{g_1e_1}(\varepsilon)}{2}(\Ket{g_1}\Bra{g_1}-\Ket{e_1}\Bra{e_1}) \\
        & +\frac{\delta_{g_2e_2}(\varepsilon)}{2}(\Ket{g_2}\Bra{g_2}-\Ket{e_2}\Bra{e_2})
    \end{split}
    \label{eq:simHamilt}
\end{equation}
where $\alpha_c\approx-\dfrac{\chi_1^2(\omega_c-\omega_{q1})}{\alpha_1^2}-\dfrac{\chi_2^2(\omega_c-\omega_{q2})}{\alpha_2^2}\approx40$ KHz is the cavity self-Kerr \cite{Elliott_2018}, which had a negligible effect, and $\delta_{ij}(\varepsilon)$ being the Stark shift of the $i\leftrightarrow j$ transition frequency induced by both the Zeno drive and the symmetric drive with an amplitude of $\varepsilon$. Including $\delta_{ij}(\varepsilon)$ in the Hamiltonian corresponds to driving the Stark shifted frequency in the experiment, as we have done. The Stark shift was calibrated beforehand by simulating a Ramsey experiment with the drives on and extracting the phase accumulation rate between all the relevant states.

\subsection*{Calibration of cavity drives}
\begin{figure}[htp!]
    \centering
    \includegraphics[width=0.5\linewidth]{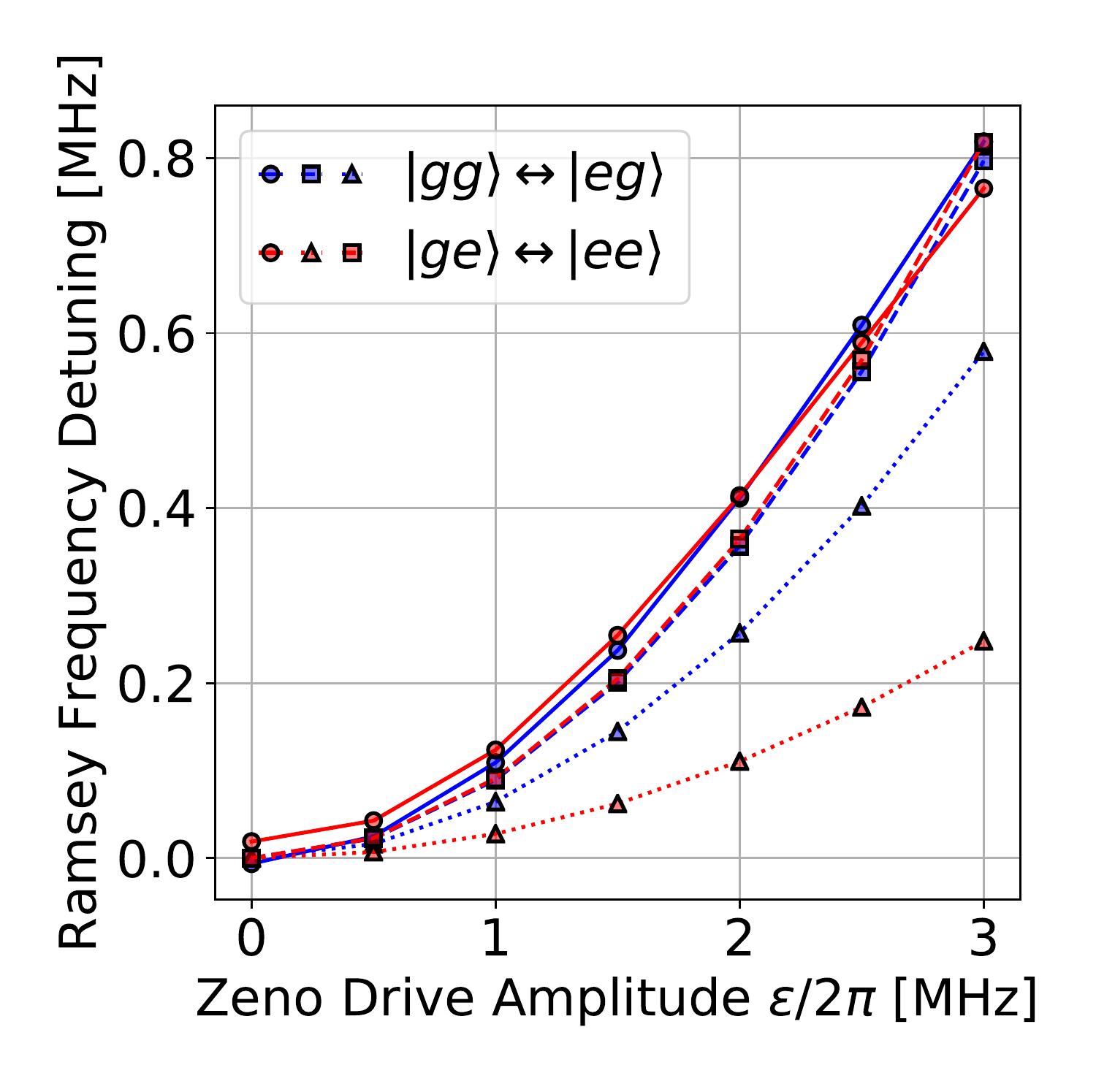}
    \caption{The mean Ramsey frequency detuning from an artificial $5$ MHz as a result of the Zeno drive and the symmetric drive as a function of their amplitude $\varepsilon$. We measured oscillations between $\Ket{g}$ and $\Ket{e}$ of the qutrit when the qubit is at $\Ket{g}$ (blue) and $\Ket{e}$ (red). Circles represent experimental results and squares represent simulated results. The additional triangles represent results of a simulation without the symmetric drive for reference. The simulated phase was linearly fit to obtain the detuning, although it showed small oscillations around the linear line. These oscillation's amplitude increased with $\varepsilon$ and were negligible for small amplitude ($\varepsilon<2.5$ MHz.} 
    \label{fig:RIPcancelation}
\end{figure}

We calibrated the amplitude of both the Zeno drive and the symmetric drive using Ramsey interferometry on the first transition of the qutrit $q_1$. We measured the detuning of the Ramsey frequency due to the Stark shift induced by each drive separately for various set voltages. By fitting the data to 
\begin{equation} \mathrm{Re}[\mu_{gg,eg}]=\dfrac{(\omega_{gg}-\omega_{eg})|\varepsilon|^2}{\Delta_{gg}\Delta_{eg}}, 
\end{equation} 
we were able to calibrate the drive amplitude $\varepsilon$ that is actually incident on the cavity. Following this calibration, we repeated the measurement with both drives applied simultaneously, once when the qubit $q_2$ was in the ground state and once when it was in the excited state (shown in Fig.~\ref{fig:RIPcancelation}). The induced Stark shift was almost equal in both cases, which means the symmetric drive does indeed cancel the entangling phase accumulation, as desired. We measured and simulated this shift for all the possible transitions and used the shifted frequencies to drive them when applying the Zeno gate.
\subsection*{Tomography details}
To perform full state tomography we measure all the operators that span the Hilbert space, which is a total of 36 operators for a qutrit and a qubit. We partially map each operator to the projector we measure $\Ket{gg}\Bra{gg}$, as was done in \cite{MattReedThesis}.
We then acquire the expectation value of each of the 36 operators. To reconstruct the density matrix we use a method of maximum likelihood estimation (MLE) as described in \cite{SINGH20163051} to find the most likely valid density matrix. If the trace of the density matrix is less than $1$, the MLE process will effectively add an appropriately scaled completely mixed state to achieve $\text{Tr}[\rho]=1$.

\subsection*{Different initial states and incorrect mapping error}
\begin{figure*}[htp!]
    \centering
    \includegraphics[width=0.57\linewidth]{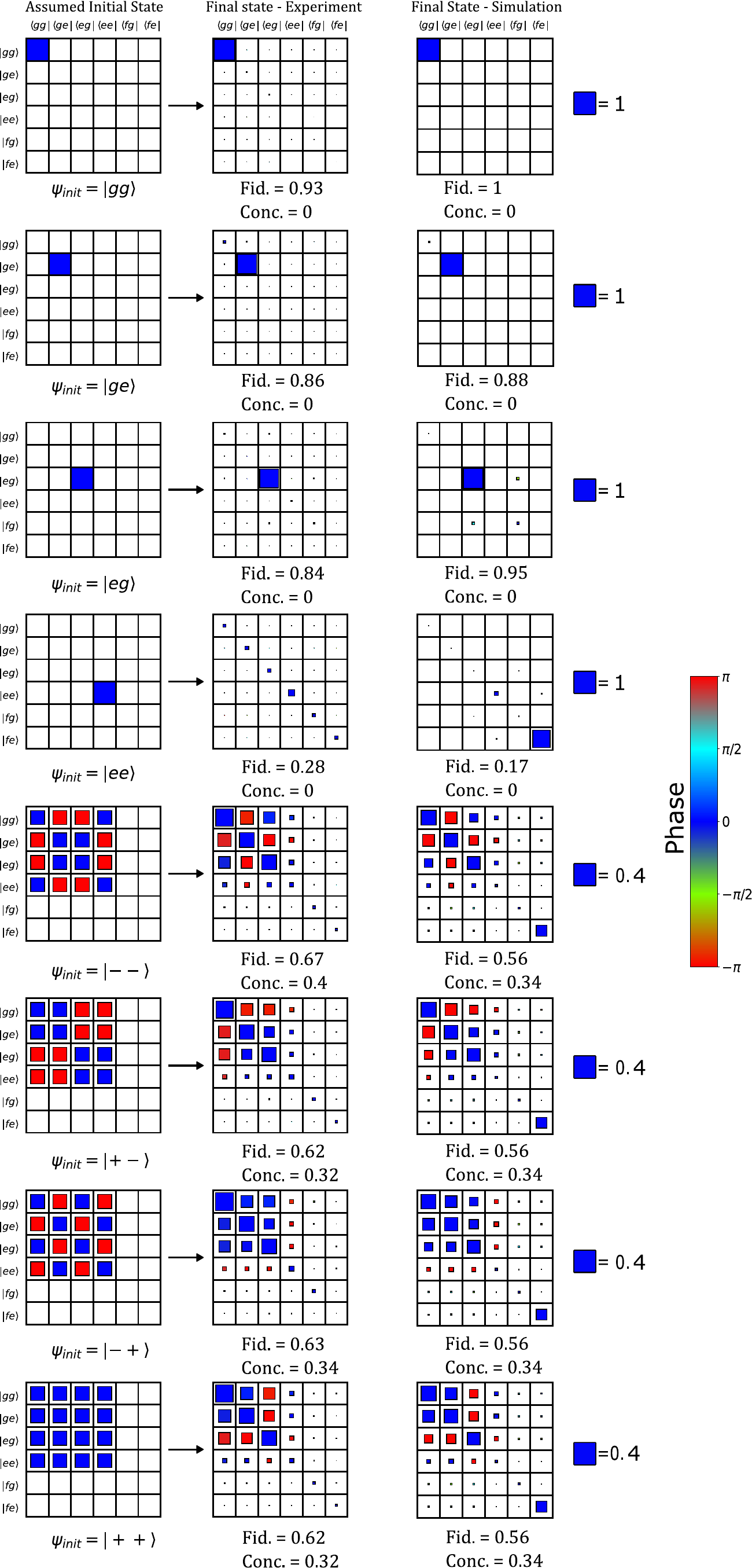}
    \caption{Experimental tomography and simulation results of Zeno gate application on different initial states. The black squares are partially filled to represent the amplitude, and the color of the filling represent the phase, according to the color bar.} 
    \label{fig:AllInits}
\end{figure*}

\begin{figure}[htp!]
    \centering
    \includegraphics[width=0.65\linewidth]{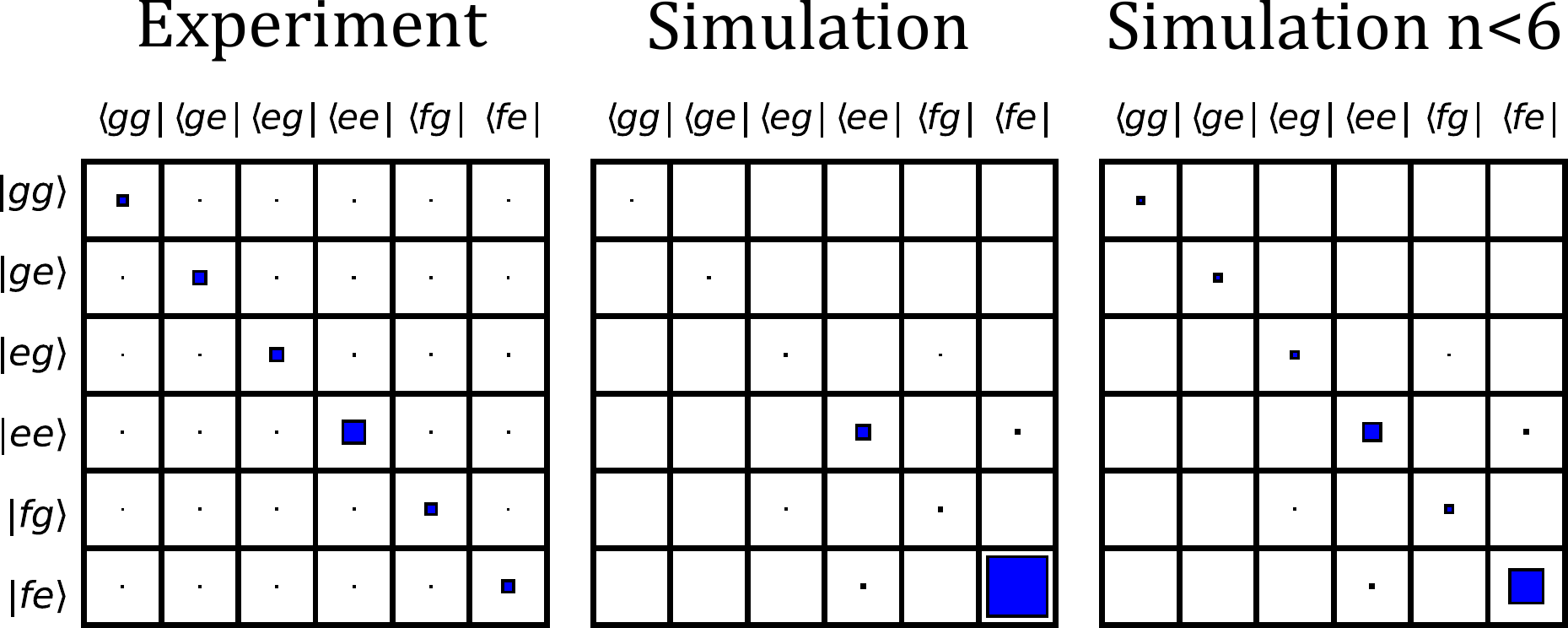}
    \caption{A comparison between experiment, simulation and truncated simulation showing only states with $n<6$ of the Zeno gate applied to an initial state $\Ket{ee}$. The density matrix from the truncated simulation was reconstructed using MLE method. Truncating states with high photon numbers is roughly similar to applying pulses with finite frequency for tomography. Indeed, we see greater overlap between the experiment and the truncated simulation.} 
    \label{fig:n-cut}
\end{figure}

To better understand the effect of the Zeno gate on our system, we applied it to different states in the computational qubit-qubit subspace, and evaluated the resulting fidelity and concurrence. The states were $\Ket{gg},\Ket{ge},\Ket{eg},\Ket{ee},\Ket{++},\Ket{-+},\Ket{+-}$ and $\Ket{--}$. The results are shown in Fig.~\ref{fig:AllInits}. The escape from the Zeno subspace error is evident in the application of the gate on $\Ket{ee}$. In addition, a considerable deviation from the simulation is clear and is due to the incorrect mapping error. The simulation shows large population in $\Ket{fe}$ but the experiment shows an almost completely mixed state. The pulses we used to control the system where Gaussian pulses with $\sigma\approx20$ ns, so that their width in Fourier space is $\Delta f\approx50$ MHz. As $\chi_f=10$ MHz, the pulse would not map states of the type $\Ket{fe}\otimes\Ket{n}$ with $n\gtrapprox5$ correctly. We used a toy model to examine this error. We reconstructed the simulated density matrix after truncating all the states with $n>5$. Fig.~\ref{fig:n-cut} shows the resulting density matrix compared with the experimental result and the full simulation. The truncated case fits better with experiment, which strengthens our assumption that this error is causing the discrepancy. However, a more accurate approach, such as including the tomography pulses in the simulation, could be used, but may not necessarily give a more significant insight.
\subsection*{Post-selection procedure}
\begin{figure}[htp!]
    \centering
    \includegraphics[width=0.5\linewidth]{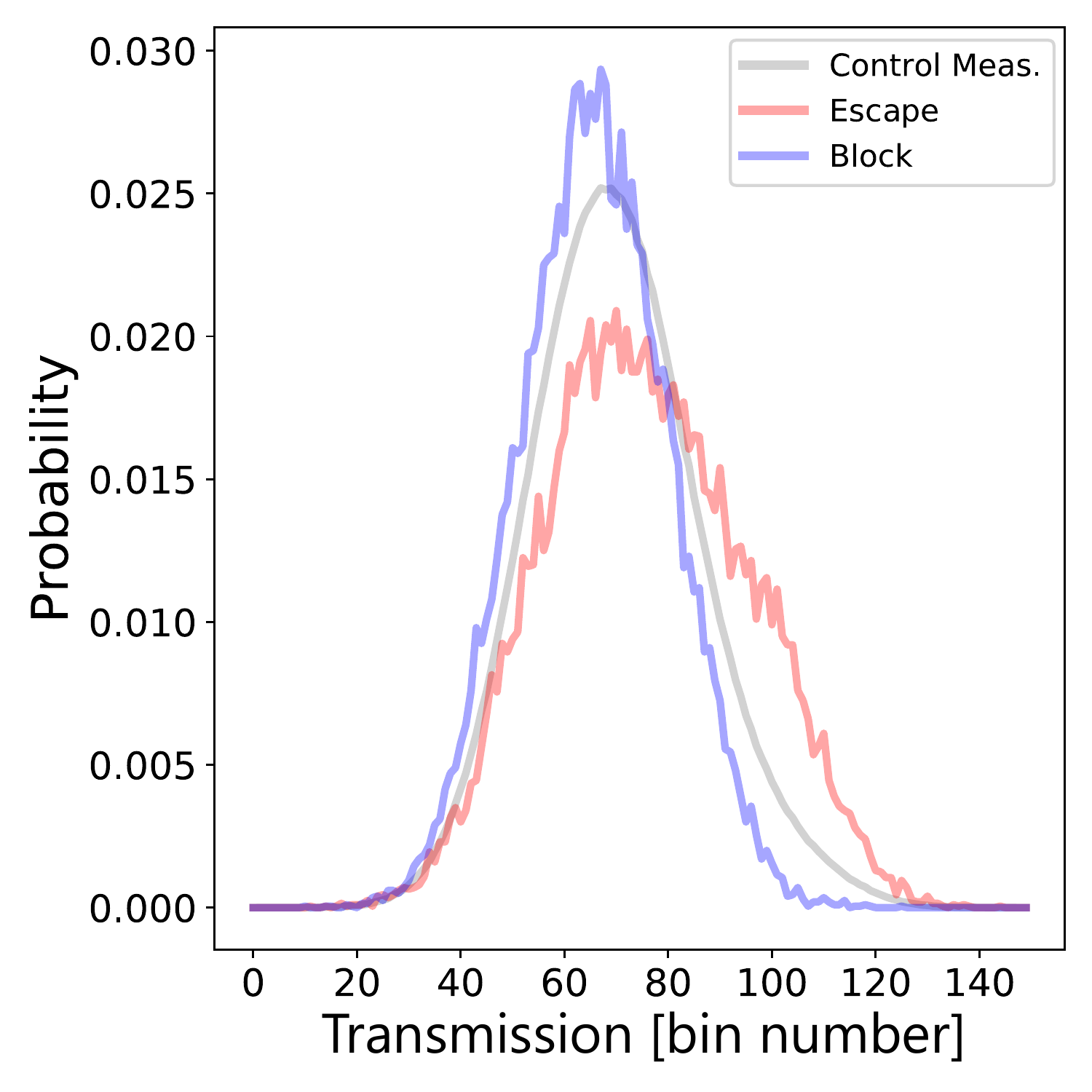}
    \caption{Histogram of the in-phase amplified transmitted signal amplitude at frequency $\omega_{fe}$. The red and blue lines are for reference, representing $\Ket{fe}$ (escape) and $\Ket{ee}$ (block), respectively. We prepared each state and then applied a readout pulse for a time that is equal to the gate time $t=\dfrac{2\pi}{\Omega_R}$. The gray line is the control measurement of the Zeno drive signal that we detected during the time of the gate. We used this control measurement for our post selection procedure.} 
    \label{fig:PostSelectionData}
\end{figure}

We used an error-detection and post-selection procedure to improve the performance of the Zeno gate (Fig.~\ref{fig:fdltyVSEpsilon}b).  High amplitude of the amplified signal that was transmitted through the cavity during the time of the gate indicates an escape to the blocked $\Ket{fe}$ state (as shown in Fig.~\ref{fig:PostSelectionData}). We post-selected our tomography data by ignoring measurements with amplitudes above a set threshold when reconstructing the density matrix. The post-selection percentage is the percentage of measurements we ignore, which was increased by decreasing the set threshold.

\subsection*{Zeno dynamics in the Markovian and non-Markovian regimes}
In this experiment we worked in the non-Markovian regime, which cannot be described by the simple form of Eq.~\ref{eq:idealMasterEq} in the main text; nevertheless our system displayed Zeno dynamics leading to entanglement, and showed quantitative agreement with numerical simulations. It would be instructive to model the system in this non-Markovian regime and see if a closed form expression connecting the effective blocking rate and the applied drives can be derived. We leave this question to future theoretical work. 

\begin{figure}[htp!]
    \centering
    \includegraphics[width=0.7\linewidth]{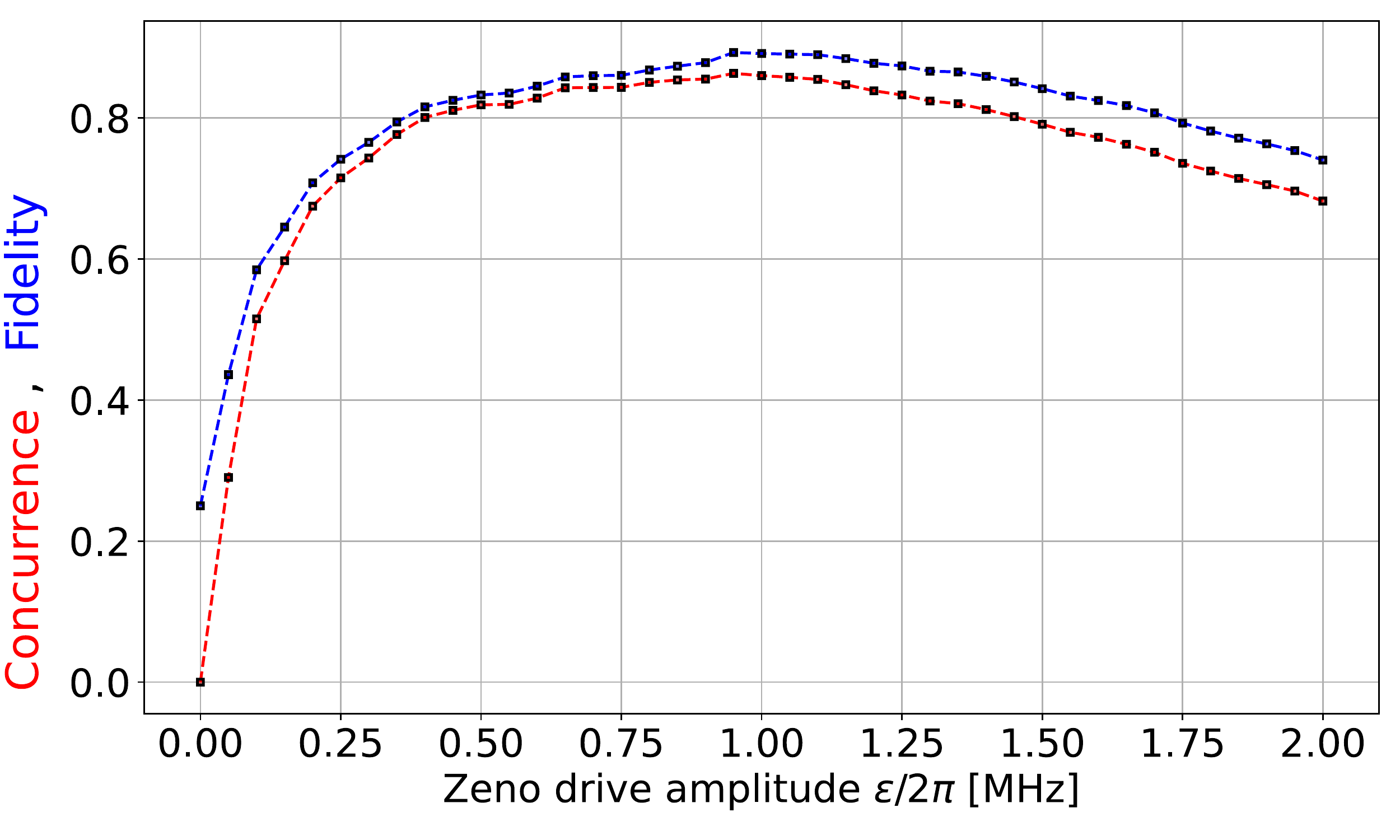}
    \caption{Results of a simulation of the Zeno gate with the same $\chi$'s and $\kappa$ as in the experimental setup but different $\Omega_R=0.1$ and infinite coherence times. Gate fidelity (blue) and concurrence (red) versus the amplitude of the Zeno drive are shown. Lines are guide to the eye. We see that in the Markovian regime the gate fidelity is increased significantly.} 
    \label{fig:epsilon sweep long time}
\end{figure}

In terms of the effect itself, the Markovian regime can yield much improved gate fidelity. The effective measurement rate can be much larger compared with the competing dephasing. This can be reached by increasing $\kappa$ and $\chi$, or slowing down the gate. For example if in our system the gate is slowed to 10$\mu$s and ignoring coherence times, fidelity of 90\% can be achieved (as shown in Fig.~\ref{fig:epsilon sweep long time}). 

\end{document}